\documentclass[a4paper, amsfonts, amssymb, amsmath, preprint, showkeys, twoside, superscriptaddress]{revtex4-1}

\usepackage{xcolor}
\usepackage[utf8]{inputenc}
\usepackage{adjustbox}
\usepackage[colorlinks,allcolors=black,citecolor=blue,urlcolor=blue]{hyperref}

\newcommand{\vk}[1]{{\color{black} #1}}
\newcommand{\highlight}[1]{{\color{black} #1}}

\DeclareUnicodeCharacter{2009}{\,} 

\usepackage[]{multibib}
\newcites{supp}{Methods References}
\bibliographystylesupp{naturemag}

\begin{document}

\title{The first-principles phase diagram of monolayer nanoconfined water}

\author{Venkat Kapil}
\email[Correspondence email address: ]{vk380@cam.ac.uk}
\affiliation{Yusuf Hamied Department of Chemistry,  University of Cambridge,  Lensfield Road,  Cambridge,  CB2 1EW, United Kingdom}

\author{Christoph Schran}
\email[Correspondence email address: ]{cs2121@cam.ac.uk}
\affiliation{Yusuf Hamied Department of Chemistry,  University of Cambridge,  Lensfield Road,  Cambridge,  CB2 1EW, United Kingdom}
\affiliation{Department  of  Physics  and  Astronomy, University College London, Gower Street, London, WC1E 6BT, United Kingdom}
\affiliation{Thomas Young Centre and London Centre for Nanotechnology,17-19 Gordon Street, London WC1H 0AH, United Kingdom}

\author{Andrea Zen}
\affiliation{Dipartimento di Fisica Ettore Pancini, Universit\`a di Napoli Federico II, Monte S. Angelo, I-80126 Napoli, Italy}
\affiliation{Department of Earth Sciences, University College London, Gower Street, London WC1E 6BT, United Kingdom}

\author{Ji Chen}
\affiliation{School of Physics, Peking University, Beijing, 100871, China}

\author{Chris J. Pickard}
\affiliation{Department of Materials Science and Metallurgy, University of Cambridge, Cambridge CB30FS, United Kingdom}
\affiliation{Advanced Institute for Materials Research, Tohoku University, Sendai 980-8577, Japan}

\author{Angelos Michaelides}
\email[Correspondence email address: ]{am452@cam.ac.uk}
\affiliation{Yusuf Hamied Department of Chemistry,  University of Cambridge,  Lensfield Road,  Cambridge,  CB2 1EW, United Kingdom}
\affiliation{Department  of  Physics  and  Astronomy, University College London, Gower Street, London, WC1E 6BT, United Kingdom}
\affiliation{Thomas Young Centre and London Centre for Nanotechnology,17-19 Gordon Street, London WC1H 0AH, United Kingdom}

\maketitle

\section*{Summary}

\noindent \highlight{Water in nanoscale cavities is ubiquitous and of central importance to everyday phenomena in geology and biology.
However, the properties of nanoscale water can be remarkably different from bulk, as shown e.g., by the anomalously low dielectric constant of water in nanochannels~\cite{fumagalli_anomalously_2018}, near frictionless water flow ~\cite{secchi_massive_2016}, or the possible existence of a square ice phase~\cite{algara-siller_square_2015}.
Such properties suggest that nanoconfined water could be engineered for technological applications in nanofluidics~\cite{kavokine_fluctuation-induced_2021}, electrolyte materials~\cite{hummer_water_2001}, and water desalination~\cite{surwade_water_2015}. 
Unfortunately, challenges in experimentally characterising water on the nanoscale and the high cost of first-principles simulations have prevented the molecular level understanding required to control the behavior of water.
Here we combine a range of computational approaches to enable a first-principles level investigation of a single layer of water within a graphene-like channel. 
}
We find that monolayer water exhibits surprisingly rich and diverse phase behavior that is highly sensitive to temperature and the van der Waals pressure acting within the nanochannel.
In addition to multiple molecular phases with melting temperatures varying non-monotonically by over 400 degrees with pressure, we predict a hexatic phase, which is an intermediate between a solid and a liquid, and a superionic phase with a high electrical conductivity exceeding that of battery materials. 
Notably, this suggests that nanoconfinement could be a promising route towards superionic behavior at easily accessible conditions.
\\

\newpage
\section*{main}

\noindent In the last decades, novel approaches have been devised to fabricate artificial hydrophobic capillaries with nanoscale dimensions~\cite{radha_molecular_2016, geim_van_2013, nigues_ultrahigh_2014}, resulting in unprecedented measurements of the properties of nanoconfined water~\cite{algara-siller_square_2015, secchi_massive_2016, fumagalli_anomalously_2018}. 
Despite these remarkable advances, the understanding of nanoconfined water is still limited, due to the difficulty in interpreting these experiments in the absence of simultaneous information on the atomistic-scale structure and dynamics of water -- not yet accessible in these experiments. 
\vk{Molecular simulations can in principle provide the required resolution but the results obtained are highly sensitive to the approach used, even for a single layer of water within graphene confinement. 
For instance, studies involving inexpensive empirical water models have predicted varied phase-behavior depending on the choice of the model~\cite{zhu_superheating_2017, li_replica_2019}.
Furthermore, their non-reactive nature makes them unsuitable for studying water at high pressures where water can dissociate and even exhibit superionic behaviour~\cite{millot_nanosecond_2019, millot_experimental_2018}.
In contrast, (more accurate) first-principles studies of nanoconfined water, using density functional theory (DFT), are limited by their large computational cost and have thus only been limited to 0\,K calculations, or a handful of finite-temperature state points~\cite{jiang_first-principles_2021}.
Thus, the absence of accurate but affordable first-principles approaches that can explore a range of temperatures and pressures has lead to a general lack of knowledge on the phase behavior of nanoconfined water, ranging from, different stable phases, melting temperatures, nature of phase-transitions, to physical properties relevant to nanotechnology.}
A clear understanding of the phases and properties of confined water as a function of thermodynamic parameters will facilitate interpretation of experiments, and provide a foundation for the rational design of improved nanotechnologies. \\

\noindent  Here, we describe the phase behavior of monolayer confined water by calculating its pressure-temperature phase diagram at first-principles accuracy.
We avoid the traditional accuracy-cost trade-off by using quantum Monte Carlo (QMC), one of the most accurate first-principles methods for molecular materials~\cite{zen_fast_2018}, to identify the most appropriate DFT functional for the systems under investigation.
We subsequently develop a machine learning potential (MLP) using the approach recently introduced in ~\cite{schran_machine_2021} to predict DFT energies and forces at a much lower cost.
\vk{This framework brings together the high accuracy of QMC and the low computational cost of MLPs, allowing us  to explore the pressure-temperature phase diagram -- including the dissociation regime. 
In doing so, we reconcile inconsistencies between previous studies, unveil new phases and properties of confined water that can be harnessed for nanotechnology, and provide guidance for future experimental exploration.}
\section*{Extensive polymorph search using an MLP}
\noindent  Water under nanoconfinement experiences a lateral pressure due to a combination of attractive van der Waals (vdW) forces pulling the sheets of the confining material and the termination of the cavities.
The exact value of this lateral pressure within graphene confinement is not known, but computational studies and experiments suggest it to be in the 0.5 to 1.5\, GPa range~\cite{algara-siller_square_2015}. 
These pressures can stabilize polymorphs of ice that are not found in bulk~\cite{corsetti_structural_2016, chen_two_2016}.
The lateral pressure, however, depends on the magnitude of the vdW forces~\cite{bjorkman_van_2012} as well as on the shape and mechanical response of the terminal edges of the cavity~\cite{zamborlini_nanobubbles_2015}, both of which are sensitive to the nature of the confining material.
Therefore, studying the phase behavior of nanoconfined water across a broad range of pressures, beyond the expected 0.5 to 1.5\,GPa regime,  is relevant for a general understanding of the properties of water in hydrophobic nanocavities. \\

\noindent Before computing the phase diagram, we identify all polymorphs stabilized by lateral pressures between 0\, and 4\,GPa.
To accurately account for the full range of interactions in nanoconfined water, while ensuring a viable computational cost to perform an extensive structure search, we use an MLP trained on first-principles energies and forces.
\vk{We use QMC benchmarks to select the revPBE0-D3 DFT functional for treating water-water interactions in confinement (see comparison in Extended Material (EM) Table~\ref{tab:lattice_energy}) , and subsequently train and validate an MLP to revPBE0-D3 across the entire phase-diagram, using an advanced active learning approach~\cite{schran_machine_2021}  (see details on training and validation in EM Fig.~\ref{fig:active_learning}).}
We treat graphene-water interactions using a uniform 2D confining potential fitted to QMC water-graphene binding energies and apply an external lateral pressure~\cite{corsetti_structural_2016, chen_evidence_2016, li_replica_2019, zhu_superheating_2017} to model the internal lateral pressure on the water molecules.  \\

\noindent Thanks to the computational efficiency of the MLP, we can perform an exhaustive structure search~\cite{pickard_ab_2011}.
We recover all previously found polymorphs~\cite{chen_two_2016, corsetti_structural_2016} in just a few hours on a desktop machine, and find a novel ``flat-rhombic" phase at high pressures \vk{(see EM Fig.~\ref{fig:ice_phases}).}
This structure is stabilized by a strong network of zig-zagging hydrogen bonds, which differs from the motifs present in flat-rhombic phases observed in empirical force-field studies.~\cite{li_replica_2019, zhu_superheating_2017}.
We attribute the differences to the rigid water models employed in these studies that do not describe the complex hydrogen bond network arising from the full spectrum of O--H distances and H--O--H angles present in our structures \vk{(see EM Fig.~\ref{fig:ice_phases}).}
To establish the relative stability of this new phase, we use enthalpies calculated with QMC at 0\, and 2\, GPa \vk{(see EM Table~\ref{tab:lattice_energy}).} 
In agreement with previous results, hexagonal ice has the lowest enthalpy at 0\,GPa~\cite{chen_evidence_2016}, however, beyond 2\,GPa we find that the flat-rhombic phase becomes the dominant form, being more stable than square ice by about 1\,kJ/mol \vk{(see EM Fig.~\ref{fig:QMC_0K_phase_diagram}).} \\

\begin{figure*}[!t]
    \centering
    \includegraphics[width=\textwidth]{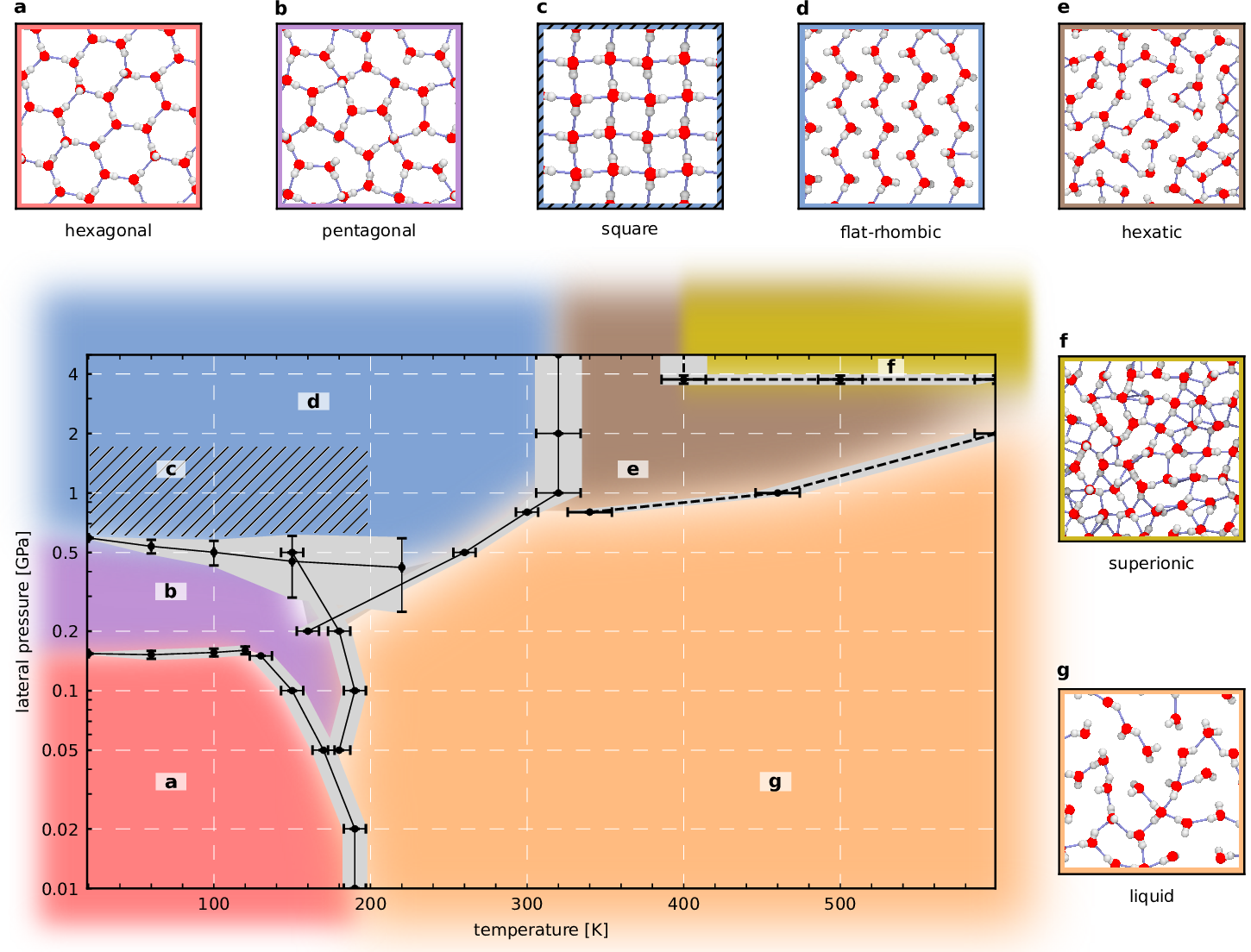}\\
    \caption{\scriptsize\textbf{Phase diagram of monolayer nanoconfined water.} The pressure-temperature phase diagram of monolayer water calculated using a machine learning potential that delivers first-principles accuracy. Solid and dashed lines indicate first-order and continuous phase transitions, respectively. Gray regions indicate the statistical uncertainty for solid-solid phase transitions, and for the other transitions, the uncertainties arising from studying a finite number of thermodynamic states. The diagonally hatched area indicates the region where \vk{ square and flat-rhombic} phases are near degenerate. Snapshots of different phases are shown with oxygen atoms in red, hydrogen atoms in gray, and hydrogen bonds depicted by blue lines.}
    \label{fig:phase_diagram}
\end{figure*}

\noindent Encouraged by the success of the MLP in exhaustively identifying monolayer polymorphs, and to find the MLP predictions within 1\,kJ/mol of the QMC references \vk{(see EM Table~\ref{tab:lattice_energy})}, we proceed to explore the pressure-temperature phase diagram.
We select a small confinement width of 5\,\AA{}, as the effects of nanoconfinement should be the largest in the narrow width regime. 
\vk{As detailed in the methods section}, we use a range of rigorous free energy methods to calculate the phase boundaries~\cite{kapil_complete_2022}.
The resulting phase diagram, shown in  Fig.~\ref{fig:phase_diagram}, is extremely rich and diverse for a deceptively simple monolayer of water.
The phase behavior is very sensitive to the lateral pressure, and thus, in the following, we discuss the phase diagram in different pressure regimes: low pressures below 0.5 GPa; intermediate pressures from 0.5 - 2.0\,GPa i.e. at the expected lateral vdW pressure within graphene confinement; and high pressures above 2.0 GPa. 
As seen from \vk{EM Tables~\ref{tab:sensitivity_potential}, EM Tables~\ref{tab:sensitivity_width},  and EM Fig.~\ref{fig:sensitivity_FSNQE}}, our predictions are robust with respect to the choice of the confining potential, small ($\sim$ 0.5 \AA{}) changes in the confinement width, as well as finite-size effects. We also show that nuclear quantum effects have only a small impact on the phase boundaries, and therefore for clarity, they are not taken into account in Fig.~\ref{fig:phase_diagram}.

\section*{Rich phase behavior at low pressures}

\noindent  Like bulk water, hexagonal ice is the most stable phase at zero pressure for monolayer water. 
As the pressure is increased, pentagonal and flat-rhombic forms become the most stable phases at around 0.15\,GPa and 0.5\,GPa, respectively.
Our QMC calculations show that the flat-rhombic phase is the most stable phase at high pressures, however, at intermediate pressures, it is only marginally more stable (within the statistical error of QMC) than the square phase. 
Thus, we have marked a narrow region on our phase diagram where the square phase could compete in thermodynamic stability with the flat-rhombic phase.\\ 

\noindent To understand the melting behavior of monolayer ice we calculate the coexistence lines of the solid and the liquid phases.
We find that hexagonal ice exhibits a low melting temperature of around 190\,K at 0\,GPa, about 100\,K lower than that of bulk water, and 50-100\,K lower than the estimates from empirical force fields~\cite{ferguson_computational_2012, zhu_superheating_2017}.
As shown in Fig. \ref{fig:phase_diagram}, we observe a triple point at around 170\,K and 0.05\,GPa  between the hexagonal, pentagonal, and liquid phases \vk{(see supplementary data (SD) video V1)}. 
As the pressure is increased, we observe a second triple point between the flat-rhombic, pentagonal, and liquid phases (see SD video V2) at around 160\,K and 0.20\,GPa.
We also note that the hexagonal, pentagonal, and flat-rhombic phases can be denser or rarer than the liquid phase, which leads to a non-monotonic dependence of the melting temperature with pressure.
This suggests that that monolayer water may expand or contract upon melting, depending on the pressure.
Thus, overall, we find that at pressures below 0.5\,GPa, monolayer ice exhibits rich polymorphism, two triple points, and a non-monotonic melting temperature with pressure.
These insights are expected to be key for the thermal or mechanical engineering of water-based nanodevices.

\begin{figure*}[!t]
    \centering
    \includegraphics[width=0.90\textwidth]{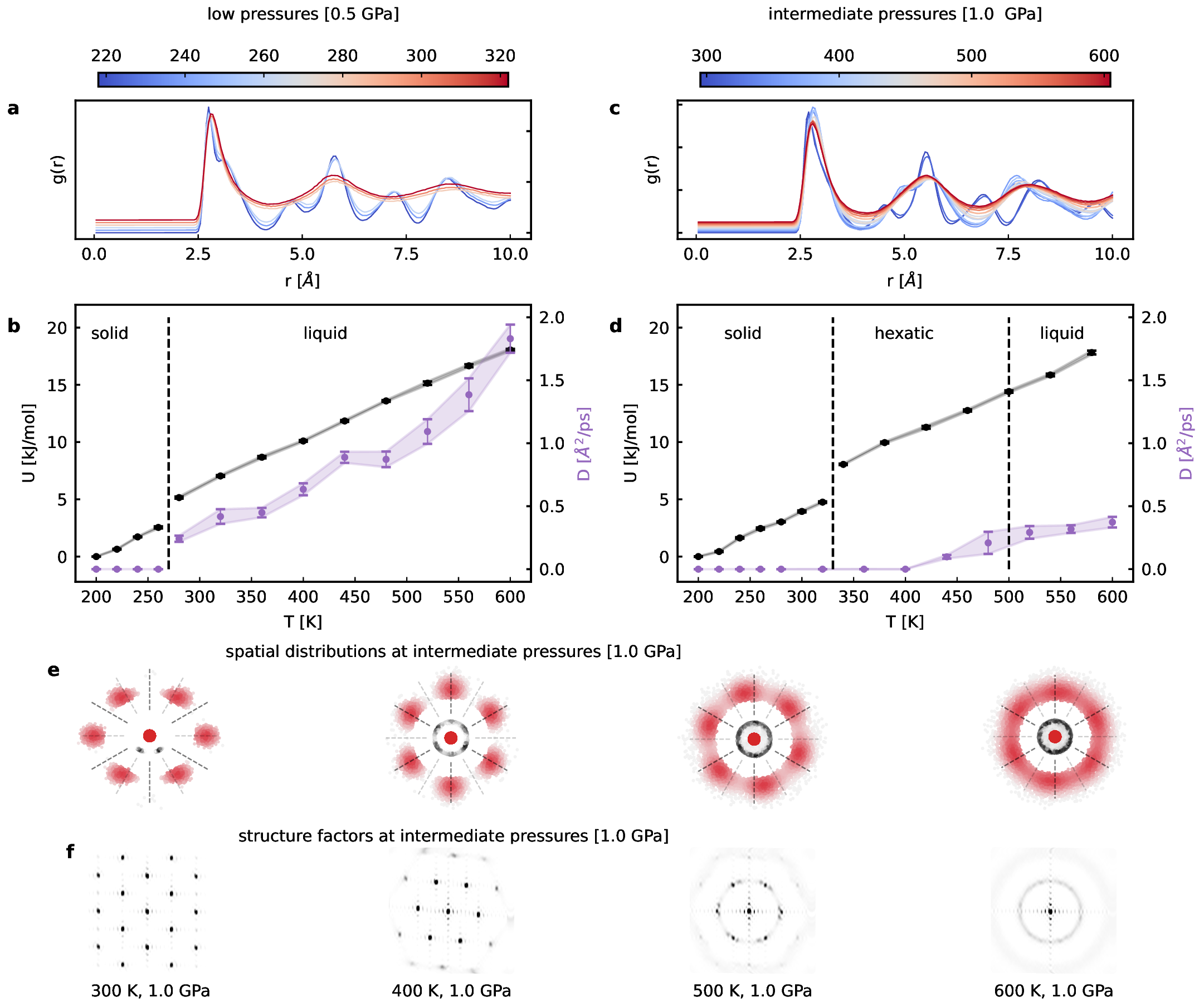}\\
    \caption{\scriptsize\textbf{Hexatic phase of monolayer water at intermediate pressures.} \textbf{(a,b)} The lateral  distribution function of oxygen atoms $g(r)$ at different temperatures and the potential energy $U$ (black) and the diffusion coefficient $D$ (violet), as a function of temperature at 0.5\,GPa, indicating a first-ordered phase transition. \textbf{(c,d)} The lateral  distribution function of oxygen atoms $g(r)$ at different temperatures and the potential energy (black) and the diffusion coefficient (violet) as a function of temperature at 1.0\,GPa, indicating a first-order phase transition to a \vk{hexatic} phase, and a continuous phase transition to the liquid phase. \textbf{(e)} The temperature dependent spatial distribution of the nearest oxygen (red) and the bonded hydrogen atoms (black) with respect to an oxygen atom, at 300\, K to 600\,K at 1\,GPa. Darker colors indicate higher spatial probabilities, and radial rulers are guides for examining 6-fold orientational symmetry -- characteristic of a hexatic phase. \textbf{(f)} The temperature dependent structure factor, indicating the loss of translational order in the hexatic phase at 400\,K, and the loss of orientational order in the liquid at 600\,K.
\label{fig:hexatic}}
\end{figure*}

\section*{Hexatic phase at intermediate pressures}
In the intermediate pressure regime, i.e. at typical vdW pressures between 0.5 and 2.0\,GPa in graphene confinement, monolayer water exhibits a phase that is neither solid nor liquid. 
As shown in Fig.~\ref{fig:hexatic}(a,b), below 0.8\,GPa, flat-rhombic ice melts via a first-order phase transition accompanied by abrupt changes in the enthalpy and the structure of water, as well as in the diffusion coefficient.
However, beyond this pressure, flat-rhombic ice undergoes a first-order phase transition to a new phase with rotating water molecules.
As evidenced in the structure-factor, shown in Fig.~\ref{fig:hexatic}(f), the new phase lacks translational order but exhibits 6 fold orientational symmetry. 
Furthermore, the angular nodes in the spatial distributions of oxygen atoms, shown in Fig.~\ref{fig:hexatic}(e), indicate long-range orientational order.
These properties suggest that the phase bears hallmarks of a hexatic phase, an intermediate phase between the solid and the liquid within the theory of phase transitions in 2D~\cite{kosterlitz_ordering_1973}  \vk{by Kosterlitz, Thouless, Halperin, Nelson and Young (KTHNY theory).} \\

\noindent \vk{The hexatic phase, at 340\,K, is characterized by} fixed but rotating water molecules as visualized from the 2D spatial distribution of hydrogen atoms \vk{(see SD video V3)}, bearing similarities with the high-pressure ``free rotor" phase of molecular hydrogen~\cite{pickard_structure_2007}.
A close look at the rotational dynamics \vk{(see SD video V3)} and the spatial distribution of atoms (see Fig. \ref{fig:hexatic}(e)), suggests that the pseudo-rotation comprises of out-of-plane motion of O--H bonds through intermediate dangling orientations, with less frequent in-plane rotations.
With an increase in temperature, the in-plane rotation becomes facile, resulting in loss of six-fold orientational symmetry and formation of an isotropic liquid as shown in Fig. \ref{fig:hexatic}(e).
The \vk{hexatic} phase melts with smooth changes in enthalpy, structure, and diffusion coefficient, indicating a second-order phase transition in agreement with the KTHNY theory. 
In short, at typical pressures experienced in graphene confinement, the melting behavior of monolayer water deviates significantly from that of bulk, \vk{exhibiting a two-step mechanism, in agreement with KTHNY theory, via a hexatic phase with ``rotating" water molecules}.  \\

\noindent Although the hallmark/s of an intermediate \vk{hexatic} phase have been discussed qualitatively in previous forcefield studies \cite{vilanova_structural_2011}, our first-principles level prediction in the intermediate pressure regime suggests that this phase can be realized experimentally in graphene confinement.
We provide further support for this by simulating water between explicit graphene sheets at the reference first-principles level \vk{(see SD video V6)}, and observing hexatic behavior in agreement with the prediction of our MLP.
Finally, to facilitate the identification of this phase we  calculate the rotational relaxation time of water molecules and the temperature-dependent vibrational spectra, which can be compared with nuclear magnetic resonance and Raman spectroscopic measurements, respectively \vk{(see EM Fig.~\ref{fig:hexatic_measurables})}. \\

\begin{figure*}[!t]
    \centering
    \includegraphics[width=\textwidth]{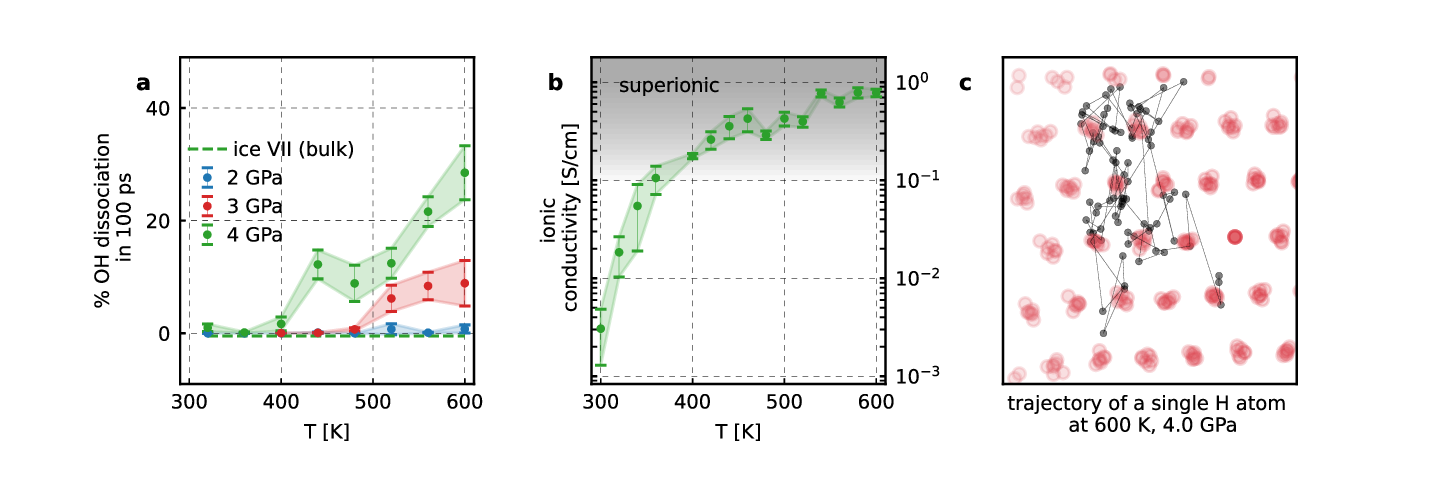}\\
    \caption{\scriptsize\textbf{Superionic behaviour at high pressures.}
    \textbf{(a)} Temperature dependence of the percentage of dissociated OH bonds in monolayer (circles) and bulk (dashed line) water in 100 ps, at 2 (blue), 3 (red) and 4 (green) \,GPa. The values for ice VII at 4\,GPa are all almost zero but have been vertically adjusted to a small negative value for clarity. \textbf{(b)} Temperature dependent ionic conductivity of monolayer water at 4\,GPa. Gray region indicates the advent of the superionic regime. \textbf{(c)} Spatial distribution of oxygen atoms (red) with respect to an arbitrarily chosen oxygen atom, and the trajectory of a single hydrogen atom (black) at 600\,K and 4\, GPa.
\label{fig:superionic}}
\end{figure*}
\section*{Superionic behavior at high pressures}

\noindent One of the key features of our water model is its ability to describe water dissociation, which becomes increasingly relevant at high pressures.
We thus explore the possibility of autoprotolysis events in monolayer water. 
We find that at pressures beyond 2 GPa and temperatures beyond 350\,K O--H bond breaking and formation events frequently occur.
To quantify O--H dissociation we analyze the average number of dissociated water molecules within 100\,ps at different pressures and temperatures. 
As shown in Fig. \ref{fig:superionic}(a), at 4 GPa and beyond 400\,K, over 10 \% of the water molecules dissociate in monolayer water within 100\,ps. 
In contrast, this percentage remains close to zero for the corresponding bulk phase (ice VII) even at the highest temperature considered.
We attribute this significantly greater propensity for dissociation to an approximate 10\% decrease in O--O distances in monolayer water compared to the bulk; it is known that the O--O distance modulates the proton transfer barrier~\cite{munoz-santiburcio_confinement-controlled_2021}.
We also note that oxygen atoms remain localized, but protons readily diffuse through the oxygen lattice (see Fig. \ref{fig:superionic}(c) and video V4).
This behavior is characteristic of the recently observed superionic phase of water~\cite{millot_experimental_2018}, believed to be found under extreme planetary conditions. 
To ascertain if we are in the superionic regime, we estimate the ionic conductivity of monolayer water as a function of temperature at 4\,GPa.
As shown in Fig. \ref{fig:superionic}(b), water ``continuously" transforms to a phase with a specific ionic conductance above 0.1\,S/cm, a commonly used threshold for characterizing a superionic conductor~\cite{kreuer_proton_1996}.
This is intriguing as bulk water exhibits superionic behavior at 56\,GPa and over 700\,K~\cite{sugimura_experimental_2012}, which is an order of magnitude greater pressure and almost twice the temperature than for monolayer water. \\

\noindent To clarify if our observations correspond to a conventional superionic phase of water in which protons diffuse within a matrix of oxygen atoms or a phase in which ionic conductivity occurs through surface hopping of protons, we performed first-principles simulations with explicit graphene sheets \vk{(see SD video V5)}. 
Although within explicit graphene, protons have the opportunity to stick to and hydrogenate the graphene sheet, these simulations confirm the results obtained with the MLP: There is facile O--H dissociation and long-range proton diffusion without the formation of C--H bonds.
This observation is consistent with the known affinities of protons with carbon and oxygen and, more importantly, it suggests that nanoconfined water can exhibit genuine superionic behavior as opposed to an interfacial layer with high ionic conductivity. \\

\noindent The enhanced propensity of autoprotolysis in confined water should be manifested at typical lateral pressures in the ``in-plane" component of the dielectric constant, that is sensitive to excursions of protons along the O--H bonds.
This property can already be measured for nanoconfinement between graphene and other hydrophobic materials~\cite{fumagalli_anomalously_2018}.
Direct observation of superionic conduction may require lateral pressures higher than those typically exerted by graphene sheets.
To this extent, it may be useful to understand how the lateral pressures can be tuned by altering the morphology of different confining materials.
Experiments already suggest that graphene sheets distorted through ``nanobubbles" can exert pressures in the tens of GPa regime~\cite{zamborlini_nanobubbles_2015} over a length scale of tens of nanometers.
We believe that nanocavities made of similarly deformed materials  could be a starting point towards `high-pressure confinement". \\

\section*{Conclusions}

\noindent We predict the pressure-temperature phase diagram of monolayer confined water rigorously at first-principles accuracy -- from low temperatures and pressures up to the dissociation regime of water.
A comprehensive description of the complex behavior of nanoconfined water across the temperature-pressure phase diagram provides new insights, a guide for future experiments, and a starting point for rational material design in the context of nanotechnology.
The existence of a two-step melting mechanism in agreement with KTHNY theory and a \vk{hexatic} phase provides new physical insight that the melting of water at the nanoscale may not necessarily be first-ordered as in bulk, and may have direct implications in the thermal engineering of nanosystems.
\vk{The evidence for increased water dissociation in nanoconfinement provides a basis for rationalizing anomalous properties of nanoconfined water, for instance, the low slippage in water flow in h-BN~\cite{secchi_massive_2016} could be understood on the basis of the enhanced propensity of O--H dissociation in nanoconfined water.}
\vk{Furthermore, the large electrical conductivity of superionic nanoconfined water could be harnessed for the development of aqueous electrolytes.}
Finally, our work suggests that nanoconfinement could be a route towards probing superionic materials at easily accessible conditions.

\newpage
\section*{Methods}
{
\noindent \highlight{To ensure that our approach remains both computationally efficient and accurate, we employ a combination of these four developments:
\begin{enumerate}
    \item An efficient implementation of QMC to select a DFT functional;
    \item An automated development and validation of the MLP to DFT energies and forces;
    \item Random structure search with the MLP to find metastable polymorphs;
    \item Rigorous Gibbs free energies as the basis for thermodynamic stabilities of polymorphs.
\end{enumerate}

\noindent Step 1 attenuates the limitations of DFT by identifying the most appropriate DFT functional for the system under study, and makes accurate first-principles energies and forces accessible for training an MLP -- at a much lower cost than high level electronic structure methods. 
Step 2 dramatically reduces the human time required to develop an MLP trained on DFT energies and forces. Step 3 ensures an exhaustive search (due to the low cost of the MLP) for metastable polymorphs of the system. Step 4 allows for a realistic modeling of experimental thermodynamic conditions and calculation of observables that can be directly compared with experiments. \\

\noindent Below, we briefly discuss the full workflow for the calculation of the phase-diagram and the computational settings, and refer the reader to the SI for all other details.}

\subsection*{Workflow}

\subsubsection{Separation of the potential energy surface}

\noindent We split the total potential energy of the system, into graphene-water interactions modeled using a uniform 2D confining potential fitted to QMC water-graphene binding energies~\citesupp{chen_two_2016}, and water-water interactions at the DFT level of theory. 
The confining potential only acts in the direction perpendicular to the plane of confinement, as in previous studies~\citesupp{chen_evidence_2016, li_replica_2019}. 
We model the lateral pressure, due to the vdW interactions pulling the graphene sheets and the termination of the edges, by applying an external lateral pressure within the $NP_{xy}T$ ensemble.

\subsubsection{Selection of an accurate DFT level for water-water interactions}

\noindent \highlight{We exploit the ``chemical accuracy" and efficiency of a recent ``fixed-node diffussion Monte Carlo" implementation of QMC~\citesupp{zen_fast_2018}}, to provide lattice energies of a few known phases of monolayer water~\citesupp{chen_two_2016} -- hexagonal, pentagonal, square, and \vk{buckled-rhombic} ice. These values are used as references to screen suitable DFT functionals for water-water interactions. From these validation tests, we select revPBE0-D3 as it exhibits one of the smallest errors in lattice energies, and is also known to provide a good description of the properties of bulk water~\citesupp{cheng_ab_2019}. 

\subsubsection{Development of the machine learning potential to DFT}

\noindent We train an MLP to reproduce revPBE0-D3 energies and forces, in order to probe new phases of monolayer confined water and their phase-behavior across a wide range of temperatures and pressures, in an accurate and efficient manner.
\vk{
In order to minimize the time spent in developing the potential, we use a recently proposed active learning scheme~\citesupp{schran_machine_2021}.
It relies on the Behler-Parrinello neural network framework~\citesupp{behler_four_2021} to form a committee neural network potential (C-NNP)~\citesupp{schran_committee_2020} for active learning purposes.
With this workflow we target various phase points in the course of different generations as detailed in Fig.~\ref{fig:active_learning}.
}
Our final potential exhibits a root mean square error of 2.4\,meV (0.2\,kJ/mol) per water molecule and 75\,meV/\AA{}  for energies and forces, respectively.
Furthermore, the MLP also reproduces the good agreement of revPBE0-D3 with QMC reference lattice energies \vk{as shown in Tab.~\ref{tab:lattice_energy} and remains accurate over the entire phase diagram, while retaining excellent performance for condensed phase properties as shown in Fig.~\ref{fig:active_learning}.}

\subsubsection{Random structure search}
\noindent We probe potential phases of monolayer ice using the random structure search approach~\citesupp{pickard_ab_2011} in combination with the MLP. Within this approach, we optimize a large set of structures generated by randomly placing water molecules within the region of confinement for unit cells with 2 to 72 water molecules, confinement widths ranging from  5 to 8 \AA{}, and lateral pressures from 0 to 4\,GPa. We recover all previously known ice polymorphs, and also find a new flat-rhombic phase stable at high pressures.

\subsubsection{Calculation of phase boundaries}
\noindent With the knowledge of stable and metastable phases, we calculate their regions of stabilities by computing their Gibbs free energies, $G(P,T)$, using a thermodynamic integration approach~\citesupp{kapil_complete_2022}.  \vk{At each pressure, we calculate the classical harmonic Helmholtz free energy at 20\,K ($A^{\text{harm}}_{20\,K}$) of a given polymorph, and perform a harmonic-to-anharmonic thermodynamic integration to estimate
\begin{equation}
G(P,T_{20\,K}) = A^{\text{harm}}_{20\,K} + \Delta A^{\text{anharm}}_{20\,K} + P V + k_{\text{B}} (20\,K) \ln p(V)_{P,20\,K},
\end{equation}
where, $V$ is the volume of the structure optimized at pressure P, and $p(V)_{P, 20\,K}$ is its probability at 20\,K and pressure P. The temperature dependent Gibbs free energy is estimated from a thermodynamic integration over temperature. The phase diagram for the polymorphs is calculated by identifying the phase with the lowest Gibbs free energy at each $P,T$. } \\

\noindent \vk{The solid to disordered phase boundary at each pressure is calculated by monitoring migration of phase boundaries in direct coexistence simulations~\citesupp{conde_determining_2013}. The initial configurations for these simulations are generated carefully in 3 steps: (i) we thermalize a $2\times1\times1$ simulation cell of a polymorph at $P,T$ (ii) we fix the oxygen atoms of one half of the cell, and melt the other half, and subsequently cool the full system to $T$ in the (constrained) NVT ensemble (iii) relax the density of the full system in the $N P_{x} T$ ensemble.} \\

\noindent \vk{To estimate the boundary of the continuous phase-transition between the hexatic-like and the liquid phase we use a criterion based on six-fold orientational symmetry. Using these approaches in combination with MLPs,  we take into account finite temperature thermodynamics at first-principles accuracy. We also check the importance quantum nuclear effects by performing path-integral molecular dynamics simulation. In all cases we calculate uncertainties due to statistical and systematic errors.}

\subsection*{Computational details}

\noindent All MLP simulations are performed using the \texttt{i-PI}~\citesupp{kapil_i-pi_2019} code with the \texttt{n2p2-LAMMPS}~\citesupp{singraber_library-based_2019} library to calculate the MLP energies and forces, and an \texttt{ASE}~\citesupp{larsen_atomic_2017} driver for the uniform confining potential.
Simulation supercells containing 144 molecules are used for all phases after carefully examining finite size effects.
Electrical conductivities are estimated using Green–Kubo theory on the basis of atomistic velocities and fixed atomistic oxidation numbers~\citesupp{grasselli_topological_2019}.
The revPBE0-D3 calculations are performed with \texttt{CP2K}~\citesupp{kuhne_cp2k_2020} with computational settings from an earlier work~\citesupp{marsalek_quantum_2017}. \\

\noindent \vk{The first-principles molecular dynamics simulations are carried out with the \texttt{CP2K} package using the revPBE-D3 and revPBE0-D3 functionals. The model consists of two layers of graphene and an embedded layer of 2D ice. The superionic and the hexatic structures were stretched slightly ($<$ 10 \%) to fit the lattice of graphene supercells. The supercell, including the vacuum slab, is $12.3 \times 8.52144 \times 20.0 ~\AA{}$. In each graphene layer, 8 evenly distributed atoms were not allowed to move along the perpendicular direction during the AIMD simulations to avoid buckling. The hexactic and superionic phase were simulated at 400\,K and 600\,K in the NVT enemble, for 50 ps at revPBE-D3 level, and for 10 ps at revPBE0-D3 level. Both sets of simulations exhibited the same qualitative behavior. } \\

\noindent The QMC calculations are performed using the \texttt{CASINO} package~\citesupp{casino} using Dirac-Fock pseudopotentials with the locality approximations~\citesupp{mitas91}, while taking into accounts errors due to finite time step and systems sizes~\citesupp{zen_fast_2018}.

\noindent \vk{More technical details of all the simulations / calculations can be found in their corresponding input files (see source data).}

\section*{Figure Legends}

\subsection*{Fig. \ref{fig:phase_diagram}} 
\noindent \textbf{Phase diagram of monolayer nanoconfined water.} The pressure-temperature phase diagram of monolayer water calculated using a machine learning potential that delivers first-principles accuracy. Solid and dashed lines indicate first-order and continuous phase transitions, respectively. Gray regions indicate the statistical uncertainty for solid-solid phase transitions, and for the other transitions, the uncertainties arising from studying a finite number of thermodynamic states. The diagonally hatched area indicates the region where \vk{square and flat-rhombic} phases are near degenerate. Snapshots of different phases are shown with oxygen atoms in red, hydrogen atoms in gray, and hydrogen bonds depicted by blue lines.

\subsection*{Fig. \ref{fig:hexatic}}
\noindent \textbf{Hexatic phase of monolayer water at intermediate pressures.} \textbf{(a,b)} The lateral  distribution function of oxygen atoms $g(r)$ at different temperatures and the potential energy $U$ (black) and the diffusion coefficient $D$ (violet), as a function of temperature at 0.5\,GPa, indicating a first-ordered phase transition. \textbf{(c,d)} The lateral  distribution function of oxygen atoms $g(r)$ at different temperatures and the potential energy (black) and the diffusion coefficient (violet) as a function of temperature at 1.0\,GPa, indicating a first-order phase transition to a hexatic phase, and a continuous phase transition to the liquid phase. \textbf{(e)} The temperature dependent spatial distribution of the nearest oxygen (red) and the bonded hydrogen atoms (black) with respect to an oxygen atom, at 300\, K to 600\,K at 1\,GPa. Darker colors indicate higher spatial probabilities, and radial rulers are guides for examining 6-fold orientational symmetry -- characteristic of a hexatic phase. \textbf{(f)} The temperature dependent structure factor, indicating the loss of translational order in the hexatic phase at 400\,K, and the loss of orientational order in the liquid at 600\,K.

\subsection*{Fig. \ref{fig:superionic}}
\noindent \textbf{Superionic behaviour at high pressures.} \textbf{(a)} Temperature dependence of the percentage of dissociated OH bonds in monolayer (circles) and bulk (dashed line) water in 100 ps, at 2 (blue), 3 (red) and 4 (green) \,GPa. The values for ice VII at 4\,GPa are all almost zero but have been vertically adjusted to a small negative value for clarity. \textbf{(b)} Temperature dependent ionic conductivity of monolayer water at 4\,GPa. Gray region indicates the advent of the superionic regime. \textbf{(c)} Spatial distribution of oxygen atoms (red) with respect to an arbitrarily chosen oxygen atom, and the trajectory of a single hydrogen atom (black) at 600\,K and 4\, GPa.

\subsection*{Fig. \ref{fig:active_learning}}
\noindent \textbf{Active learning workflow of the machine learning potential across the full phase diagram.} \textbf{(a)} As depicted in the schematic, the main idea behind the machine learning potential is the combination of multiple neural network potentials (NNPs) in a ``committee model"~\cite{schran_committee_2020}, where the committee members are separately trained by random subsampling of the  training set. Here we use a committee of 8 neural network potentials. The committee average provides more accurate predictions than the individual NNPs, and the committee disagreement, the standard deviation across the committee, is an estimate of the error of the model. To construct a training set of such a model in an automated and data-driven way, new configurations with the highest disagreement can be added to the training set. This approach is known as ``query by committee" (QbC). \textbf{(b)} As shown in the schematic, the development of MLPs for various phase points is performed across different ``generations", such that in each generation new thermodynamic state points are targeted to yield a MLP which is used to sample new candidate structures for the next generation. We used the bulk water potential from ~Ref.~\citenum{schran_committee_2020}, trained on 814 configurations of liquid water, different ice phases, and the water-vacuum interface, all including nuclear quantum effects. In the next generation we added 521 new structures of classical and path integral $NPT$ simulations of monolayer, bilayer water, bulk water and hexagonal ice (100 - 400\,K and 0 - 1\,GPa), ice VII and ice VIII (100 - 400\,K and 0.8 - 10\,GPa), and two sets of monolayer and bilayer ice structures from Ref. ~\citenum{chen_two_2016}. In the final generation we added 207 structures from classical and path integral $NPT$  temperature and pressure ramps of monolayer water (100 - 400\,K and 0 - 5\,GPa). We obtain an energy and force root mean square error (RMSE) for the training set of 2.4\,meV per H$_2$O (or 0.2 kJ/mol) and 75.4\,meV/\AA{}, respectively. \textbf{(c)} Force (top) and energy (bottom) root mean square error (RMSE) of an independent validation set covering the explored phase diagram of mono-layer confined water. The largest force RMSE for this validation set is with 100 mev/\AA{}, suggesting that the model remains accurate and robust across a wide range of temperature and pressure. The new model also keeps its excellent performance for the original condensed phase conditions, as noted by a predicted density of 0.93\,kg/L and a melting temperature of 270 $\pm$ 5 \,K at ambient pressure, in excellent agreement with the reference functional~\cite{cheng_ab_2019}.

\subsection*{Fig. \ref{fig:ice_phases}}
\noindent \textbf{(b)} Top and side view of the monolayer pentagonal phase. \textbf{(c)} Top and side view of the monolayer flat-rhombic phase. \textbf{(d)} Top and side view of the monolayer square phase. Lattice constants  are labeled along cell vectors. Insets show histograms of O--H distances and H--O--H  angles for the optimized structures following the approach from Ref.~\citenum{chen_two_2016}

\subsection*{Fig. \ref{fig:QMC_0K_phase_diagram}}
\noindent  \textbf{Zero temperature (classical) phase diagram at quantum Monte Carlo level.} The pressure dependence of the lattice enthalpies of the pentagonal (red) and square (pink) phases relative to the flat-rhombic (green) phase calculated at QMC level. Error bars indicate stochastic QMC error.

\subsection*{Fig. \ref{fig:sensitivity_FSNQE}}
\noindent\textbf{Sensitivity analysis of coexistence lines with respect to finite size and quantum nuclear effects.} \textbf{(a)} The temperature dependence of the potential energy for simulations initialized with monolayer hexagonal, pentagonal and flat-rhombic ice phases simulations at 0 pressure, 0.5\,GPa and 1.0\,GPa respectively. The number of water molecules in the simulation cell are indicated in the legend, with larger (smaller) marker sizes corresponding to larger (smaller) system sizes. The potential energy of the curves have been translated vertically to make it easier to discern the temperature of melting (indicated by a vertical dashed line) for different systems sizes. Shaded regions indicate statistical uncertainty. \textbf{(b)} The temperature dependence of the potential energy for simulations initialized with monolayer hexagonal, pentagonal and flat-rhombic ice phases at 0 pressure, 0.5\,GPa and 1.0\,GPa respectively, with nuclear quantum effects. Shaded regions indicate statistical uncertainty and the melting temperature is indicated by a vertical dashed line. \textbf{(c)} The structure factor of monolayer confined water in the hexatic phase (400\,K) and at the transition between the hexatic and the liquid phase (500\,K), without nuclear quantum effects. \textbf{(d)} The structure factor of monolayer confined water in the hexatic phase (400\,K) and at the transition between the hexatic and the liquid phase (500\,K), with nuclear quantum effects.

\subsection*{Fig. \ref{fig:hexatic_measurables}}
\noindent \textbf{Rotational dynamics of monolayer hexatic water.} \textbf{(a)} 1st order orientational autocorrelation function of the hexatic phase at 1\,GPa from 340 to 400\,K calculated as $C_1(t) = \langle P_{1} [v(0)\cdot v(t)] \rangle>$ where, v(t) is the orientation of the O--H bond at time t, and $P_{1}$ is the Legendre polynomial of degree one. \textbf{(b)}  Temperature dependence of the 1st order orientational relaxation time, computed as the time integral of $C_1(t)$, for the O--H molecular axis of the hexatic phase at 1\,GPa. The relaxation time at 340\,K is around seven times smaller than that of room temperature liquid water, indicating facile rotation of water molecules in the hexatic-like phase. \textbf{(c)} The vibrational density of states across the flat-rhombic (up to 320\,K) to the hexatic-like phase transition (beyond 340\,K) as calculated by the Fourier transform of the velocity autocorrelation function. The flat-rhombic phase is characterized by two distinct vibrational bands for hydrogen bonded and dangling O–H bonds, and clear features for the librational modes. On the other hand, the rotations in the hexatic-like phase merge the two stretching bands into a doublet, and smear out the fine structure of the librational band.
\section*{acknowledgements}
\noindent VK acknowledges funding from the Swiss National Science Foundation (SNSF) under Project $\text{P2ELP2}\_\text{191678}$ and the Ernest Oppenheimer Fund,  allocation of CPU hours by CSCS under Project ID s1000, and support from Churchill College, University of Cambridge. CS acknowledges financial support from the Alexander von Humboldt-Stiftung. AZ acknowledges financial support from the Leverhulme Trust, grant number RPG-2020-038. Calculations were also performed on the Cambridge Service for Data Driven Discovery (CSD3) operated by the University of Cambridge Research Computing Service (www.csd3.cam.ac.uk), provided by Dell EMC and Intel using Tier-2 funding from the Engineering and Physical Sciences Research Council (capital grant EP/P020259/1), and DiRAC funding from the Science and Technology Facilities Council (www.dirac.ac.uk). JC acknowledges funding from the National Natural Science Foundation of China under Grant No. 11974024, and the Strategic Priority Research Program of Chinese Academy of Sciences under Grant No. XDB33000000.
Furthermore, we are grateful to the UK Materials and Molecular Modelling Hub for computational resources, which is partially funded by Engineering and Physical Sciences Research Council (EPSRC) (grants EP/P020194/1 and EP/T022213/1).
We thank Daan Frenkel for insightful discussions, David M Wilkins for providing the code to study the angular orientation of water, Fabian L. Thiemann for help with hydrogen bond analysis, Michael Davies for help with visualization and graphics, and Pavan Ravindra for a critical reading of the manuscript. 
\section*{Author Information}

\subsection*{Affiliations}
\subsubsection*{Yusuf Hamied Department of Chemistry,  University of Cambridge,  Lensfield Road,  Cambridge,  CB2 1EW, United Kingdom}
Venkat Kapil, Christoph Schran, Angelos Michaelides

\subsubsection*{Thomas Young Centre and London Centre for Nanotechnology,17-19 Gordon Street, London WC1H 0AH, United Kingdom}
Christoph Schran, Angelos Michaelides

\subsubsection*{Department  of  Physics  and  Astronomy, University College London, Gower Street, London, WC1E 6BT, United Kingdom}
Christoph Schran, Angelos Michaelides

\subsubsection*{Dipartimento di Fisica Ettore Pancini, Universit\`a di Napoli Federico II, Monte S. Angelo, I-80126 Napoli, Italy}
Andrea Zen

\subsubsection*{Department of Earth Sciences, University College London, Gower Street, London WC1E 6BT, United Kingdom}
Andrea Zen

\subsubsection*{School of Physics, Peking University, Beijing, 100871, China}
Ji Chen

\subsubsection*{Department of Materials Science and Metallurgy, University of Cambridge, Cambridge CB30FS, United Kingdom}
Chris J. Pickard

\subsubsection*{Advanced Institute for Materials Research, Tohoku University, Sendai 980-8577, Japan}
Chris J. Pickard

\subsubsection*{Churchill College, University of Cambridge,  Storey's Way,  Cambridge,  CB3 ODS, United Kingdom}
Venkat Kapil

\subsection*{Contributions}
\noindent V.K. and A.M. conceived the study; V.K and C.S. developed the MLP; V.K., J.C, and C.J.P. implemented and performed structure search using the RSS scheme with the MLP; VK performed MLP simulations; J.C. performed first-principles simulations; A.Z. performed QMC calculations; V.K., C.S, and AM analyzed the data, designed the figures, and  wrote the paper. All authors discussed the results and commented on the
manuscript. 

\subsection*{Corresponding Authors}
\noindent Correspondence to Venkat Kapil [\href{vk380@cam.ac.uk}{vk380@cam.ac.uk}] or Christoph Schran [\href{cs2121@cam.ac.uk}{cs2121@cam.ac.uk}] or Angelos Michaelides [\href{am452@cam.ac.uk}{am452@cam.ac.uk}]
\section*{Ethics declarations}

\subsection*{Competing interests}
The authors declare no competing interests.

\section*{Data and code availability}
\noindent  The scripts, codes, and initial structures required to reproduce the key findings of this work are available at \\

\noindent \href{https://github.com/venkatkapil24/data-monolayer-confined-water-phase-diagram/}{https://github.com/venkatkapil24/data-monolayer-confined-water-phase-diagram/}

\section*{Supplementary information}

\subsection*{Supplementary Data 1}
\noindent Videos of the hexagonal-pentagonal-liquid coexistence (V1), the rhombic-pentagonal-liquid coexistence (V2), the hexatic (V3) and superionic phases (V4), and first-principles simulation of the superionic (V5) and hexatic phases (V6) with explicit graphene.

\section*{Source Data}

The source data for all the figures is made available at \\

\noindent \href{https://github.com/venkatkapil24/data-monolayer-confined-water-phase-diagram/}{https://github.com/venkatkapil24/data-monolayer-confined-water-phase-diagram/}
\newpage

\section*{Extended Data}

\begin{table}[!ht]
\scriptsize
\begin{tabular}{c | c | c | c | c | c | c }
 \hline
 & \multicolumn{6}{c}{Lattice Enthalpy} \\
 Polymorph & \multicolumn{6}{c}{[kJ/mol]} \\
 \hline
  & \multicolumn{3}{c|}{At 0\,GPa} & \multicolumn{3}{c}{At 2\,GPa} \\
 \hline
  & DFT & NNP & DMC & DFT & NNP & DMC \\
 [1.0ex] 
 \hline
 \hline
 Hexagonal & 0 & 0 & 0 & - & -  & - \\ 
 \hline
 Pentagonal & 0.5 & 0.2 & 0.4 & 4.8 & 4.5 & 3.3  \\
 \hline
 Flat-Rhombic & 2.3 & 1.3 & 2.2 & 0 & 0 & 0  \\ [1ex] 
 \hline
 Square & 3.0 & 3.1 & 1.5 &  3.0 & 1.8 & 1.0 \\
 \hline
\end{tabular}
\caption{\scriptsize \label{tab:lattice_energy} \textbf{Relative lattice energies of variable-cell optimized monolayer ice structures} Relative lattice enthalpies of different polymorphs of monolayer confined water calculated using revPBE0-D3, the corresponding MLP developed in this work, and quantum Monte Carlo (QMC) -- the reference level of theory used to select the DFT functional. QMC values have been taken from Ref.~\citenum{chen_evidence_2016} for all structures, except for the newly found flat-rhombic phase which was computed as part of this study using exactly the same computational settings as in Ref.~\citenum{chen_evidence_2016}. The dashed entries (-) suggest that the structure is not stable at the indicated pressure.}
\end{table}

\newpage
\begin{figure*}[!ht]
    \centering
    \includegraphics[width=\textwidth]{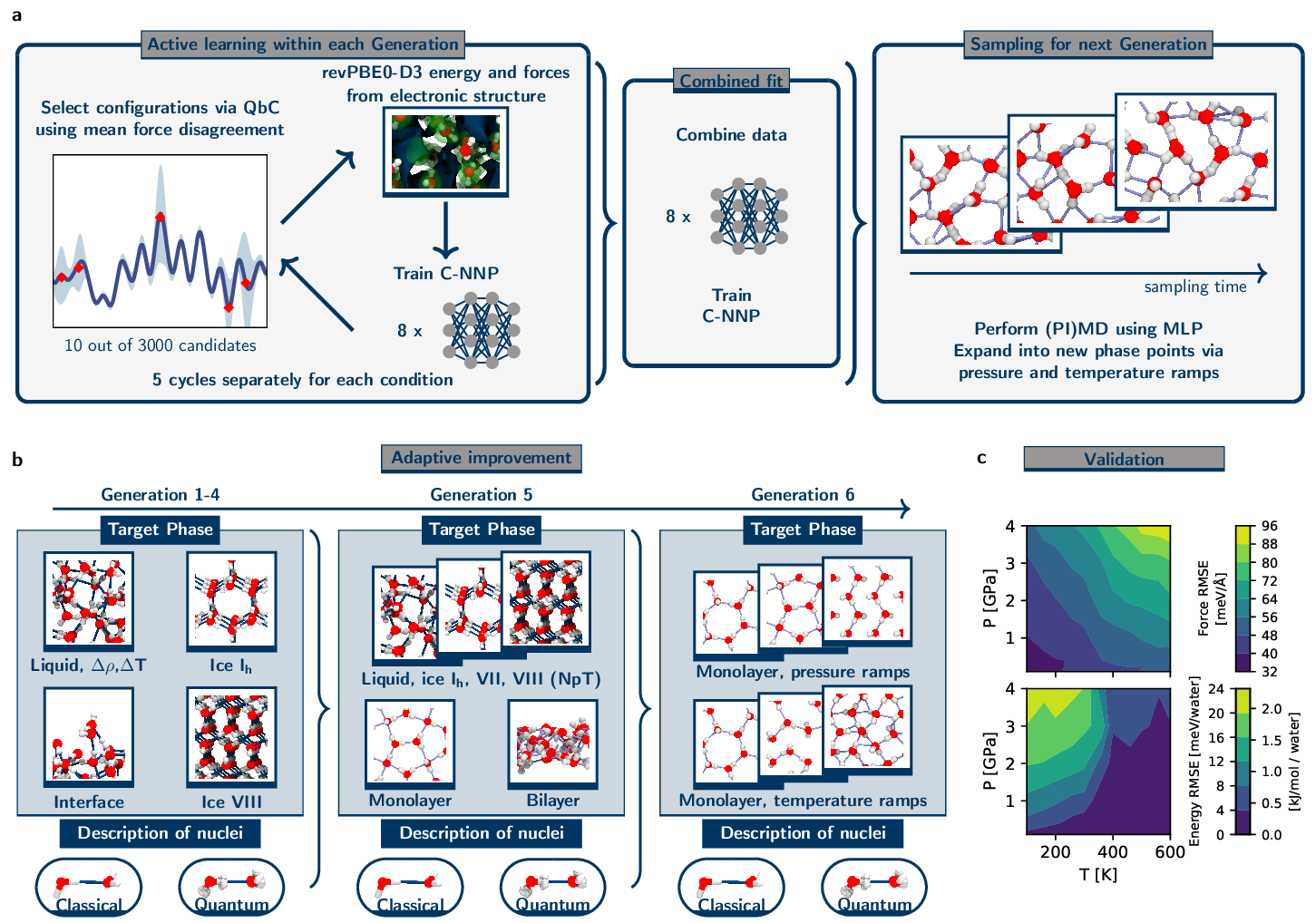}\\
    \caption{\scriptsize \label{fig:active_learning} \textbf{Active learning workflow of the machine learning potential across the full phase diagram.} \textbf{(a)} As depicted in the schematic, the main idea behind the machine learning potential is the combination of multiple neural network potentials (NNPs) in a ``committee model"~\cite{schran_committee_2020}, where the committee members are separately trained by random subsampling of the  training set. Here we use a committee of 8 neural network potentials. The committee average provides more accurate predictions than the individual NNPs, and the committee disagreement, the standard deviation across the committee, is an estimate of the error of the model. To construct a training set of such a model in an automated and data-driven way, new configurations with the highest disagreement can be added to the training set. This approach is known as ``query by committee" (QbC). \textbf{(b)} As shown in the schematic, the development of MLPs for various phase points is performed across different ``generations", such that in each generation new thermodynamic state points are targeted to yield a MLP which is used to sample new candidate structures for the next generation. We used the bulk water potential from ~Ref.~\citenum{schran_committee_2020}, trained on 814 configurations of liquid water, different ice phases, and the water-vacuum interface, all including nuclear quantum effects. In the next generation we added 521 new structures of classical and path integral $NPT$ simulations of monolayer, bilayer water, bulk water and hexagonal ice (100 - 400\,K and 0 - 1\,GPa), ice VII and ice VIII (100 - 400\,K and 0.8 - 10\,GPa), and two sets of monolayer and bilayer ice structures from Ref. ~\citenum{chen_two_2016}. In the final generation we added 207 structures from classical and path integral $NPT$  temperature and pressure ramps of monolayer water (100 - 400\,K and 0 - 5\,GPa). We obtain an energy and force root mean square error (RMSE) for the training set of 2.4\,meV per H$_2$O (or 0.2 kJ/mol) and 75.4\,meV/\AA{}, respectively. \textbf{(c)} Force (top) and energy (bottom) root mean square error (RMSE) of an independent validation set covering the explored phase diagram of mono-layer confined water. The largest force RMSE for this validation set is with 100 mev/\AA{}, suggesting that the model remains accurate and robust across a wide range of temperature and pressure. The new model also keeps its excellent performance for the original condensed phase conditions, as noted by a predicted density of 0.93\,kg/L and a melting temperature of 270 $\pm$ 5 \,K at ambient pressure, in excellent agreement with the reference functional~\cite{cheng_ab_2019}.}
\end{figure*}

\newpage
\newpage
\begin{figure*}[!ht]
    \centering
    \includegraphics[width=\textwidth]{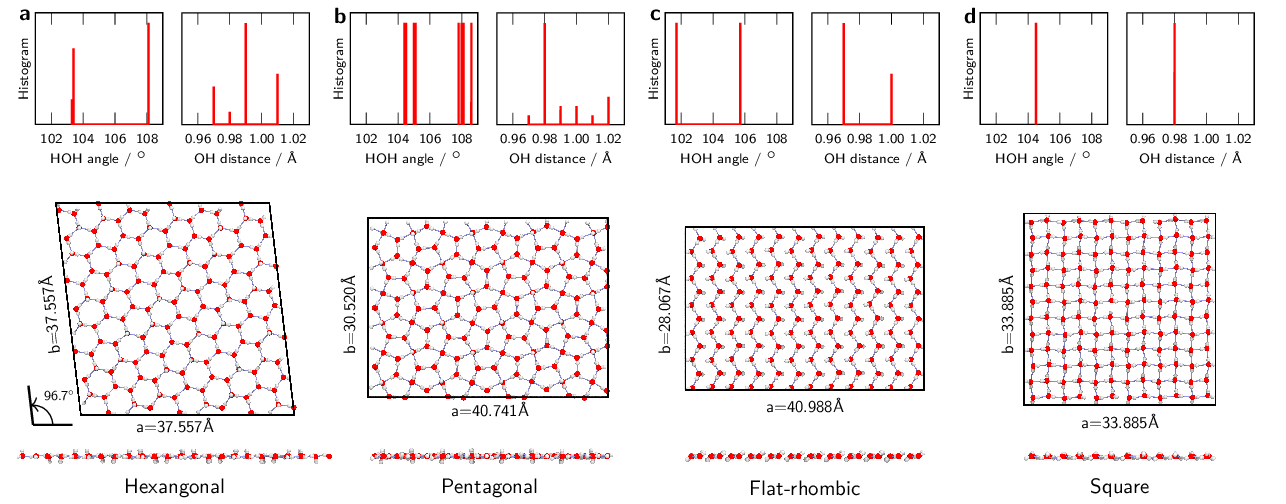}\\
    \caption{\scriptsize \label{fig:ice_phases} \textbf{Monolayer ice phases.} \textbf{(a)} Top and side view of the monolayer hexagonal phase.\textbf{(b)} Top and side view of the monolayer pentagonal phase. \textbf{(c)} Top and side view of the monolayer flat-rhombic phase. \textbf{(d)} Top and side view of the monolayer square phase. Lattice constants  are labeled along cell vectors. Insets show histograms of O--H distances and H--O--H  angles for the optimized structures following the approach from Ref.~\citenum{chen_two_2016}.}
\end{figure*}

\newpage

\begin{figure*}[!ht]
    \centering
    \includegraphics[width=\textwidth]{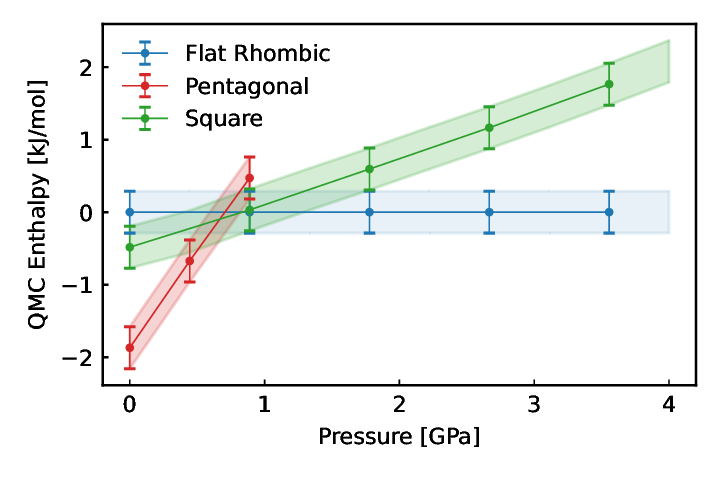}\\
    \caption{\scriptsize \textbf{Zero temperature (classical) phase diagram at quantum Monte Carlo level.} The pressure dependence of the lattice enthalpies of the pentagonal (red) and square (pink) phases relative to the flat-rhombic (green) phase calculated at QMC level. Error bars indicate stochastic QMC error.}
    \label{fig:QMC_0K_phase_diagram}
\end{figure*}

\newpage 
\begin{table}[!h]
\scriptsize
\begin{tabular}{c | c | c | c | c | c | c } 
 \hline
 & \multicolumn{6}{c}{Change in transition pressure } \\
  & \multicolumn{6}{c}{[GPa]} \\
 \hline
  & \multicolumn{3}{c|}{H - P} & \multicolumn{3}{c}{P - R} \\
 \hline
 Potential & 20 K & 60 K & 120 K & 20 K & 100 K & 200 K \\
 [1.0ex] 
 \hline
 \hline
 0 Leg & 0.000 & 0.000 & 0.000 & 0.008 & 0.006  & 0.001 \\ 
 \hline
 1 Leg & 0.003 & 0.001 & 0.010 & 0.022 & 0.165 & 0.030  \\
 \hline
 2 Leg & 0.001 & 0.001 & 0.005 & 0.116 & 0.087 & 0.016  \\ [1ex] 
 \hline
\end{tabular}
\caption{\scriptsize\label{tab:sensitivity_potential} \textbf{Sensitivity analysis of coexistence lines with respect to the parameters of the confinement potential.} Change in the transition pressure between the hexagonal (H), pentagonal (P) and the flat-rhombic (R) phases, arising from changing the parameters of the Morse potential used in this work to those of Morse potentials fitted to the QMC interaction energies of a (i) 0-leg, (ii) 1-leg, and (iii) 2-leg water molecule, taken from Ref. \citenum{brandenburg_physisorption_2019}. The confining potential used in this work is a Morse potential fitted to interaction energies of all the three types of water configurations. The changes in the transition pressures were calculated using thermodynamic perturbation theory. We note that these changes within statistical errors of the coexistence lines of the predicted phase diagram and suggest that a confinement that explicitly incorporate C--H interactions should lead to the same qualitative phase diagram.}
\end{table}

\newpage
\begin{table}[!ht]
\scriptsize
\begin{tabular}{c | c | c | c | c | c | c } 
 \hline
 & \multicolumn{6}{c}{Change in transition pressure } \\
  & \multicolumn{6}{c}{[GPa]} \\
 \hline
  & \multicolumn{3}{c|}{H - P} & \multicolumn{3}{c}{P - R} \\
 \hline
 Change in Width [\AA{}] & 20 K & 60 K & 120 K & 20 K & 100 K & 200 K \\
 [1.0ex] 
 \hline
 \hline
 0.1 & 0.000 & 0.000 & 0.003 & 0.045 & 0.034 & 0.011 \\ 
 \hline
 0.2  & 0.001 & 0.001 & 0.006 & 0.090 & 0.067 & 0.021  \\
 \hline
 0.5  & 0.003 & 0.002 & 0.015 & 0.226 & 0.168 & 0.053  \\ [1ex] 
 \hline
\end{tabular}
\caption{\scriptsize \label{tab:sensitivity_width} \textbf{Sensitivity analysis of coexistence lines with respect to the confinement width.}  Change in the transition pressure between the hexagonal (H), pentagonal (P) and the flat-rhombic (R) phases at given temperatures due to small change in the confinement width calculated using thermodynamic perturbation theory. These suggest that the predicted phase diagram is robust with respect to small changes in the confinement width.}
\end{table}

\newpage
\begin{figure*}[!ht]
    \centering
    \includegraphics[width=\textwidth]{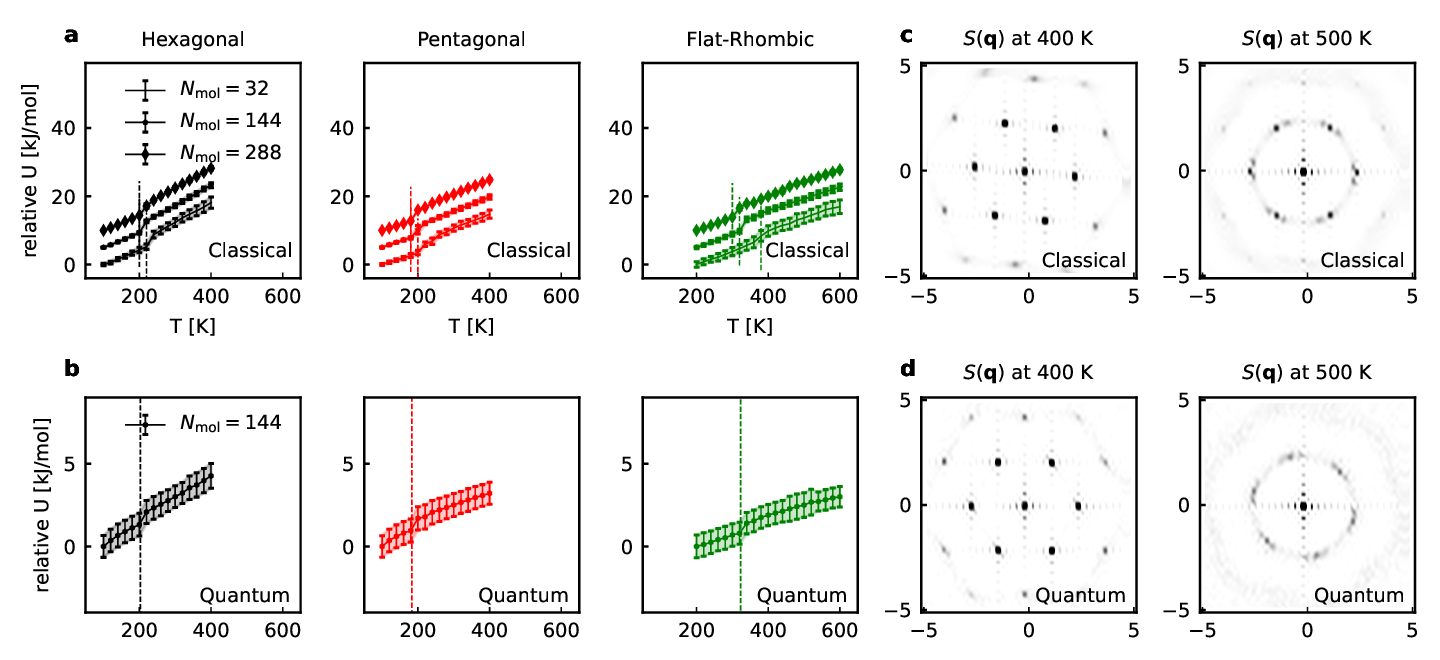}
    \caption{\scriptsize \label{fig:sensitivity_FSNQE}\textbf{Sensitivity analysis of coexistence lines with respect to finite size and quantum nuclear effects.} \textbf{(a)} The temperature dependence of the potential energy for simulations initialized with monolayer hexagonal, pentagonal and flat-rhombic ice phases simulations at 0 pressure, 0.5\,GPa and 1.0\,GPa respectively. The number of water molecules in the simulation cell are indicated in the legend, with larger (smaller) marker sizes corresponding to larger (smaller) system sizes. The potential energy of the curves have been translated vertically to make it easier to discern the temperature of melting (indicated by a vertical dashed line) for different systems sizes. Shaded regions indicate statistical uncertainty. \textbf{(b)} The temperature dependence of the potential energy for simulations initialized with monolayer hexagonal, pentagonal and flat-rhombic ice phases at 0 pressure, 0.5\,GPa and 1.0\,GPa respectively, with nuclear quantum effects. Shaded regions indicate statistical uncertainty and the melting temperature is indicated by a vertical dashed line. \textbf{(c)} The structure factor of monolayer confined water in the hexatic phase (400\,K) and at the transition between the hexatic and the liquid phase (500\,K), without nuclear quantum effects. \textbf{(d)} The structure factor of monolayer confined water in the hexatic phase (400\,K) and at the transition between the hexatic and the liquid phase (500\,K), with nuclear quantum effects.}
\end{figure*}

\newpage
\begin{figure*}[!ht]
    \centering
    \includegraphics[width=1.0\textwidth]{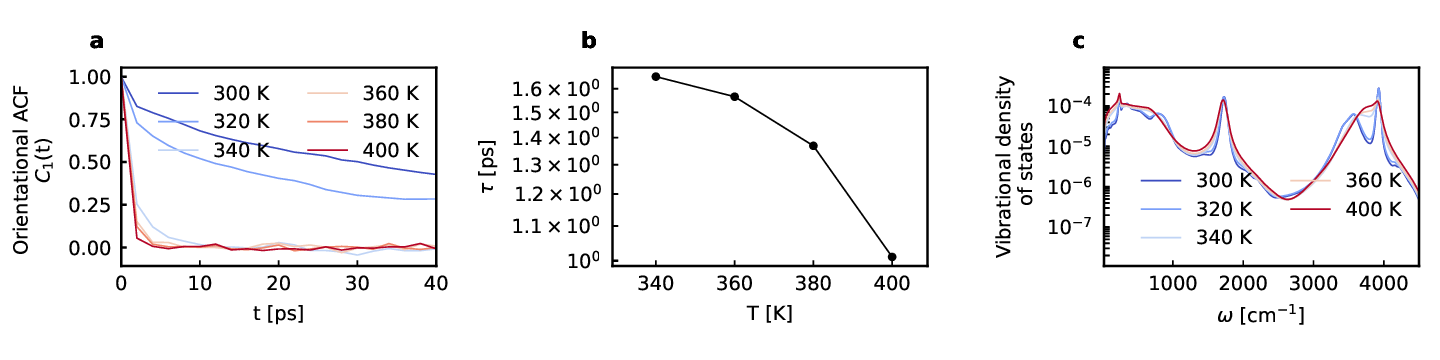}
    \caption{\scriptsize \textbf{Rotational dynamics of monolayer hexatic water.} \textbf{(a)} 1st order orientational autocorrelation function of the hexatic phase at 1\,GPa from 340 to 400\,K calculated as $C_1(t) = \langle P_{1} [v(0)\cdot v(t)] \rangle>$ where, v(t) is the orientation of the O--H bond at time t, and $P_{1}$ is the Legendre polynomial of degree one. \textbf{(b)}  Temperature dependence of the 1st order orientational relaxation time, computed as the time integral of $C_1(t)$, for the O--H molecular axis of the hexatic phase at 1\,GPa. The relaxation time at 340\,K is around seven times smaller than that of room temperature liquid water, indicating facile rotation of water molecules in the hexatic-like phase. \textbf{(c)} The vibrational density of states across the flat-rhombic (up to 320\,K) to the hexatic-like phase transition (beyond 340\,K) as calculated by the Fourier transform of the velocity autocorrelation function. The flat-rhombic phase is characterized by two distinct vibrational bands for hydrogen bonded and dangling O–H bonds, and clear features for the librational modes. On the other hand, the rotations in the hexatic-like phase merge the two stretching bands into a doublet, and smear out the fine structure of the librational band.
    \label{fig:hexatic_measurables} }
\end{figure*}

\end{document}